\documentclass[12pt]{article}
\usepackage{float} 
\usepackage{amsmath}
\usepackage{slashed}
\usepackage{amsfonts}   
\usepackage{amssymb}

\begin{document}
\date{}
\title{{\bf{\Large Multispin magnons from Spin-Matrix strings on $ AdS_5 \times S^5 $}}}
\author{
 {\bf {\normalsize Dibakar Roychowdhury}$
$\thanks{E-mail:  dibakarphys@gmail.com, dibakar.roychowdhury@ph.iitr.ac.in}}\\
 {\normalsize  Department of Physics, Indian Institute of Technology Roorkee,}\\
  {\normalsize Roorkee 247667, Uttarakhand, India}
\\[0.3cm]
}

\maketitle
\begin{abstract}
The present Letter derives multispin nonrelativistic magnon spectrum considering various near BPS corners within $ SU(1,2|3) $ Spin-Matrix theory (SMT) limit of strings on $ AdS_5 \times S^5 $. In particular, we focus on some typical rotating string solutions those correspond to two spin as well as three spin configurations in the bulk and identify the associated nonrelativistic magnon like excitations in these models. 
\end{abstract}
\section{Overview and Motivation}
The conjectured duality between the quantum mechanical/ Spin-Matrix theory (SMT) limit \cite{Harmark:2006di}-\cite{Harmark:2014mpa} of $ \mathcal{N}=4 $ SYM and the nonrelativistic strings \cite{Gomis:2000bd}-\cite{Gomis:2005pg} propagating in $ AdS_5 \times S^5 $ \cite{Harmark:2017rpg}-\cite{Harmark:2018cdl} has opened up a new era in the context of nonrelativstic gauge/string duality. Nonrelativistic string sigma models on curved manifolds seem to have two parallel formulations. One of these formulations is based on taking a large $c$ limit of relativistic string theory on string Newton-Cartan (SNC) geometry \cite{Bergshoeff:2018yvt}-\cite{Bergshoeff:2019pij}. In the second formulation, the nonrelativistic sigma model is obtained using the method of \emph{null reduction} on a (relativistic) Lorentzian manifold \cite{Harmark:2017rpg}-\cite{Harmark:2018cdl}. As an end result of this null reduction procedure, one finally arrives at Polyakov/Nambu-Goto (NG) action for nonrelativistic strings propagating on a curved manifold that has a topology of torsional Newton-Cartan (TNC) geometry times a compact (periodic) direction along which the string has a non zero winding \cite{Grosvenor:2017dfs}-\cite{Roychowdhury:2020kma}. The resulting sigma model is nonrelativistic from the perspective of the target space geometry. However, interesting physics emerges after taking a second nonrelativistic (large $c$) limit on the worldsheet degrees of freedom. This is the limit which seems to provide a dual (stringy) description \cite{Harmark:2017rpg}-\cite{Harmark:2018cdl} for the near BPS sector in $\mathcal{N}=4$ SYM \cite{Harmark:2014mpa}. The purpose of this paper is to further zoom in into this sector for a better understanding of the holographic correspondence in a nonrelativistic set up. In particular, our goal would be to explore $SU(1,2|3)$ SMT limit of sigma models in $AdS_5 \times S^5$ \cite{Harmark:2017rpg}-\cite{Harmark:2018cdl} and establish its connection with various decoupling limits \cite{Harmark:2006di}-\cite{Harmark:2014mpa} in $\mathcal{N}=4$ SYM. 

One of the major successes of the celebrated $ AdS_5/CFT_4 $ correspondence is the precise matching of the operator spectrum using both gauge theory and the string theory degrees of freedom. On the gauge theory side of the duality, one typically looks for single trace operators $ \sim tr (\Phi_X^J) $ ($ \Phi_X = \phi_1 +i \phi_2 $) with large R-charge ($ J \gg 1 $) and large conformal dimension ($ \Delta \gg 1 $) such that the difference $ \Delta -J =0 $. In the presence of finite number of other fields ($ \Phi_Z= \phi_3+ i \phi_4 $), these operators/states may be interpreted as long spin chains with added \emph{impurities} which are identified as elementary excitations associated with the spin chain known as magnons. Using SUSY, the dispersion relation (for any given values of the t'Hooft coupling ($ \lambda $)) associated with these single spin magnons have been obtained in \cite{Beisert:2005tm}. In the limit of strong coupling ($ \lambda \gg 1 $), these magnon states precisely match with the Hofman-Maldacena limit \cite{Hofman:2006xt} of semiclassical string states propagating in $ AdS_5 \times S^5 $.

The single spin solutions of \cite{Beisert:2005tm} were further generalised by taking into account the possibilities of forming magnon bound states in $ \mathcal{N}=4 $ SYM \cite{Dorey:2006dq}. On the dual string theory side of the correspondence, these bound states would correspond to the rotating two spin folded string configurations on $ R\times S^3 $ \cite{Chen:2006gea}-\cite{Bobev:2006fg}. The folded and circular two spin solutions in $ AdS_5 \times S^5 $ were further generalised in \cite{Minahan:2006bd} and a special infinite spin limit (in which the one loop correction to the classical string energy vanishes) was discussed. An intimate connection between these multispin string solutions \cite{Ryang:2006yq} and the Neumann-Rosochatius (NR) integrable models was further investigated by authors in \cite{Kruczenski:2006pk}. 

Keeping the spirit of the discussion as alluded to above, it is worthwhile to mention that a similar analysis in the SMT limit of magnons in $ \mathcal{N}=4 $ SYM and in particular unfolding its connection with nonrelativistic rotating strings propagating on $ U(1) $ Galielan geometries is still lacking in the literature. The purpose of the present paper is therefore to address some of these issues considering $ SU(1,2|3) $ SMT limit of strings on $ AdS_5 \times S^5 $ \cite{Harmark:2018cdl} and thereby providing a concrete map between the nonrelativistic stringy spectra and the corresponding quantum mechanical limit \cite{Harmark:2014mpa} of giant magnon dispersion relations in $ \mathcal{N}=4 $ SYM. In other words, our goal is to obtain nonrelativistic \emph{multispin} magnon bound states and compare it with the corresponding near BPS limits of $\mathcal{N}=4$ SYM.

To check this claim explicitly, we construct magnon bound states corresponding to two spin as well as three spin nonrelativistic strings propagating on $ U(1) $ Galilean manifolds. In case of two spin solutions, we provide an interpretation of string results from the perspective of dual quantum mechanical degrees of freedom. The key ingradient behind the above matching of spectra turns out to be the identification of the deficit angle ($ \Delta\varphi_m $) (between the end points of the string soliton) with the nonrelativistic magnon momenta ($ \frac{p_m}{c} \ll 1$) along the spin chain \cite{Roychowdhury:2020kma}. Nevertheless, a similar interpretation for the three spin solution remains to be understood.

\section{SMT strings and nonrelativistic magnons}
To start with, we consider the Polyakov action for relativistic strings
\begin{eqnarray}
S_P = \frac{\sqrt{\lambda}}{4 \pi}\int d^2 \sigma \sqrt{-\gamma}\gamma^{\alpha \beta}G_{MN}\partial_{\alpha}X^{M}\partial_{\beta}X^N~;~M,N=1,\cdots , (d+2)
\end{eqnarray}
 over a $ (d+2) $ dimensional Lorentz invariant manifold
\begin{eqnarray}
ds^2 = 2 \tau (d \mathfrak{u}-\mathfrak{m})+\mathfrak{h}_{\mu \nu}dx^{\mu}dx^{\nu}
\label{ee2}
\end{eqnarray}
associated with a null isometry direction $ X^{\mathfrak{u}}=\mathfrak{u} $. Here, $ \gamma^{\alpha \beta} $ is the 2d world-sheet metric.

Notice that, $ \tau =\tau_{\mu}dx^{\mu} $ and $ \mathfrak{m}=\mathfrak{m}_{\mu}dx^{\mu} $ introduced above in (\ref{ee2}) are the one forms together with a symmetric rank $ d $ tensor $ \mathfrak{h}_{\mu \nu} $. Finally, $ x^{\mu} $ define a set of coordinates over a $ (d+1) $ dimensional manifold that is identified as torsinal Newton Cartan (TNC) geometry.

Given (\ref{ee2}), an equivalent description for the sigma model turns out to be,
\begin{eqnarray}
S = -\frac{\sqrt{\lambda}}{4 \pi}\int d^2 \sigma (\sqrt{-\gamma}\gamma^{\alpha\beta}\bar{\mathfrak{h}}_{\alpha \beta}+(\sqrt{-\gamma}\gamma^{\alpha \beta}\tau_{\alpha}-\epsilon^{\alpha \beta}\partial_{\alpha}\eta)\mathcal{A}_{\beta})
\label{e3}
\end{eqnarray}
where,
\begin{eqnarray}
\bar{\mathfrak{h}}_{\alpha \beta}=(\mathfrak{h}_{\mu \nu}-\mathfrak{m}_{\mu}\tau_{\nu}-\mathfrak{m}_{\nu}\tau_{\mu})\partial_{\alpha}x^{\mu}\partial_{\beta}x^{\nu}.
\end{eqnarray}
Moreover, here $ \mathcal{A}_{\alpha}=\partial_{\alpha}\mathfrak{u} $ is the world-sheet one form together with the fact that $ \eta (=X^{v})$ is the world-sheet scalar which corresponds to a compact direction ($ v $) associated with the target space along which the string wraps.

Finally, following the so called \emph{null reduction} procedure \cite{Harmark:2017rpg}-\cite{Harmark:2018cdl} the above action (\ref{e3}) eventually boils down into a sigma model that corresponds to closed strings propagating over $ (d+1) $ dimensional torsinal Newton Cartan (TNC) geometry (characterized by $ \tau_{\mu} $, $ \mathfrak{m}_{\mu} $ and $ \mathfrak{h}_{\mu \nu} $) times a compact direction ($ \eta $),
\begin{eqnarray}
\label{e5}
S_{NG}=\frac{\sqrt{\lambda}}{4\pi}\int d^2 \sigma \left(-\epsilon^{\alpha \beta}\mathfrak{m}_{\alpha}\partial_{\beta}\eta +\frac{\epsilon^{\alpha \alpha'}\epsilon^{\beta \beta'}(\partial_{\alpha'}\eta \partial_{\beta'}\eta -\tau_{\alpha'}\tau_{\beta'})}{2\epsilon^{\gamma \gamma'}\tau_{\gamma}\partial_{\gamma'}\eta}\mathfrak{h}_{\alpha \beta}\right) .
\end{eqnarray}

Notice that, the above sigma model (\ref{e5}) is nonrelativistic from the perspective of the the TNC target spacetime \cite{Harmark:2018cdl}. However, one can consider a second scaling limit
\begin{eqnarray}
\label{e6}
c\rightarrow \infty ~;~\lambda = \frac{\mathfrak{g}}{c^2}~~;~~\tau_{\alpha}=c^2 \tilde{\tau}_{\alpha}~~;~~\eta =c \tilde{\eta}~~;~~\mathfrak{m}_{\alpha}=\tilde{\mathfrak{m}}_{\alpha}~~;\mathfrak{h}_{\alpha \beta}=\tilde{\mathfrak{h}}_{\alpha \beta}
\end{eqnarray}
for TNC strings (\ref{e5}) that eventually leads towards a nonrelativistic world-sheet theory. These are Spin-Matrix Theory (SMT) strings propagating over $ U(1) $-Galilean geometries\footnote{In case of $ AdS_5 \times S^5 $ strings, the SMT sigma model (\ref{e5}) is conjectured to be dual to the strong coupling dynamics in the near BPS corners of $ \mathcal{N}=4 $ SYM \cite{Harmark:2014mpa}. The near BPS limit in $ \mathcal{N}=4 $ is defined as, $\lim_{\lambda \rightarrow 0}\frac{H -Q}{\lambda} =$ fixed \cite{Harmark:2017rpg}-\cite{Harmark:2018cdl} where $ Q=S+J $ stands for the sum of the Cartan generators of the group $ SO(2,4) \times SO(6) $. Notice that for fixed $ \mathfrak{g} $ and $ c \rightarrow \infty $ the scaling limit (\ref{e6}) precisely realizes the decoupling limit $ \lambda \rightarrow 0 $ of $ \mathcal{N}=4 $ SYM while $ \mathfrak{g} $ being the coupling constant of the dual SMT theory. Here, $ H $ is the Hamiltonian of the dual Spin-Matrix theory which should be identified with the energy ($ E_{NR} $) of the SMT string at strong coupling. Similarly, the other charges ($ Q $) at strong coupling could be estimated by studying the isometries associated with the nonrelativistic string sigma model (\ref{e2}).}.

Following (\ref{e6}), we arrive at the Nambu-Goto (NG) action of the following form,
\begin{eqnarray}
\mathcal{S}_{NG}=-\frac{\sqrt{\mathfrak{g}}}{2\pi}\int d^2 \sigma \mathcal{L}_{NG}~;~\sigma^{\alpha}=\lbrace \sigma^0 , \sigma^1 \rbrace
\end{eqnarray}
where the NG Lagrangian density could be formally expressed as\footnote{For simplicity henceforth we ignore tildes.}  \cite{Harmark:2017rpg}-\cite{Harmark:2018cdl},
\begin{eqnarray}
\mathcal{L}_{NG}=\epsilon^{\alpha \beta}\mathfrak{m}_{\alpha}\partial_{\beta}\eta +\frac{\epsilon^{\alpha \alpha'}\epsilon^{\beta \beta'}\tau_{\alpha'}\tau_{\beta'}}{2 \epsilon^{\gamma \gamma'}\tau_{\gamma}\partial_{\gamma'}\eta}\mathfrak{h}_{\alpha \beta}.
\label{e2}
\end{eqnarray}

In case of $ AdS_5 \times S^5 $ strings, the world-sheet one forms are given  by \cite{Harmark:2018cdl},
\begin{eqnarray}
\tau_{\alpha}=\cosh^2\rho \partial_{\alpha}t
\end{eqnarray}
\begin{eqnarray}
\mathfrak{m}_{\alpha}=-\tanh^2\rho \left( \partial_{\alpha}\chi +\frac{1}{2}\cos\psi \partial_{\alpha}\varphi\right)~~~~~~~~~~~~~~~ \nonumber\\
+\cosh^{-2}\rho\left(\frac{1}{2}\cos\theta_2 \sin^2\theta_1 \partial_{\alpha}\phi_1 -\left( \frac{1}{3}-\frac{1}{2}\sin^2\theta_1\right)\partial_{\alpha}\phi_2 \right), 
\end{eqnarray}
together with the two form,
\begin{eqnarray}
\mathfrak{h}_{\alpha \beta}= \tanh^2\rho  \partial_{\alpha}\chi \partial_{\beta}\chi +\partial_{\alpha}\rho \partial_{\beta}\rho +\frac{1}{4}\sinh^2 \rho(\partial_{\alpha}\psi \partial_{\beta}\psi +\sin^2\psi \partial_{\alpha}\varphi \partial_{\beta}\varphi)~~~~~~~~~~~\nonumber\\
+ \tanh^2\rho \partial_{\alpha}\chi \left(\frac{1}{2}\cos\psi \partial_{\beta}\varphi +\frac{1}{2}\cos\theta_2 \sin^2\theta_1 \partial_{\beta}\phi_1 -\left( \frac{1}{3}-\frac{1}{2}\sin^2\theta_1\right)\partial_{\beta}\phi_2 \right)
+(\alpha\leftrightarrow \beta)\nonumber\\
+\frac{\tanh^2\rho}{4}\cos^2\psi \partial_{\alpha}\varphi \partial_{\beta}\varphi +\partial_{\alpha}\theta_1 \partial_{\beta}\theta_1 +\frac{1}{4}\sin^2\theta_1 (\partial_{\alpha}\theta_2 \partial_{\beta}\theta_2 +\sin^2\theta_2 \partial_{\alpha}\phi_1 \partial_{\beta}\phi_1)\nonumber\\
+\frac{1}{4}\sin^2 \theta_1 \cos^2\theta_1 (\partial_{\alpha}\phi_2 \partial_{\beta}\phi_2 + \cos\theta_2 (\partial_{\alpha}\phi_1 \partial_{\beta}\phi_2 +\partial_{\beta}\phi_1 \partial_{\alpha}\phi_2 )+ \cos^2\theta_2 \partial_{\alpha}\phi_1 \partial_{\beta}\phi_1)\nonumber\\
+\tanh^2\rho \left( \frac{\cos\psi}{2}\left(\frac{1}{2}\cos\theta_2 \sin^2\theta_1 \partial_{\alpha}\phi_1 -\left( \frac{1}{3}-\frac{1}{2}\sin^2\theta_1\right)\partial_{\alpha}\phi_2 \right) \partial_{\beta}\varphi  +(\alpha\leftrightarrow \beta)\right) \nonumber\\
+\tanh^2\rho \left(\frac{1}{2}\cos\theta_2 \sin^2\theta_1 \partial_{\alpha}\phi_1 -\left( \frac{1}{3}-\frac{1}{2}\sin^2\theta_1\right)\partial_{\alpha}\phi_2 \right) \nonumber\\
\times \left(\frac{1}{2}\cos\theta_2 \sin^2\theta_1 \partial_{\beta}\phi_1 -\left( \frac{1}{3}-\frac{1}{2}\sin^2\theta_1\right)\partial_{\beta}\phi_2 \right).
\end{eqnarray}

Here, the coordinates $ \psi $ and $ \varphi $ belong to the three sphere $ S^3 \subset AdS_5$.  On the other hand, the remaining coordinates $ \lbrace \theta_1, \theta_2, \phi_1, \phi_2\rbrace $ belong to the five sphere ($ S^5 $) which is a part of the full 10D SUGRA background.
\subsection{Two spin magnons}
We consider classical strings those are located near the center of $AdS_5$ as well as propagating in $ R \times S^3 $. We choose to work with the string embedding of the following form,
\begin{eqnarray}
t = \kappa \sigma^0 ~;~ \eta = \sigma^1  ~;~ \theta_1 = \theta (\sigma^1) ~;~ \theta_2 =0 ~;~\phi_1 =\omega_{1} \sigma^0 + \sigma^1 ~;~\phi_2 = \omega_{2} \sigma^0 + \sigma^1.
\end{eqnarray}

The resulting sigma model Lagrangian turns out to be,
\begin{eqnarray}
\mathcal{L}_{NG}=\frac{1}{2}(\omega_1 + \omega_2)\sin^2\theta -\frac{\omega_2}{3}+\frac{\kappa}{2}(\theta'^2 +\sin^2\theta \cos^2\theta)
\end{eqnarray}
which yields the equation of motion of the following form,
\begin{eqnarray}
\theta'' -\left(\frac{\omega_T}{\kappa}+1 \right) \sin\theta \cos\theta +2\sin^3\theta \cos\theta =0
\label{e23}
\end{eqnarray}
where, $ \omega_T = \omega_1 + \omega_2 $ is the total spin of the configuration.

Integrating the above equation (\ref{e23}) once we find,
\begin{eqnarray}
\theta' (\sigma^1)=\sqrt{(\alpha^2 - \gamma^2 \sin^2\theta)(\beta^2 \sin^2\theta -\mathcal{C}^2)}
\label{E9}
\end{eqnarray}
which is subjected to the constraint,
\begin{eqnarray}
\alpha^2 \beta^2 +\gamma^2 \mathcal{C}^2 &=& 1+\frac{\omega_T}{\kappa}\\
\beta^2 \gamma^2 &=&1
\end{eqnarray}
where $ \mathcal{C} $ is the constant of integration.

Introducing the limits of the $ \theta $ integral,
\begin{eqnarray}
\theta_{min}=\arcsin\left(\frac{\mathcal{C}}{\beta} \right) ~;~ \theta_{max}=\arcsin\left(\frac{\mathcal{\alpha}}{\gamma} \right)
\end{eqnarray}
one could further re-express (\ref{E9}) in terms of its roots as
\begin{eqnarray}
\theta' (\sigma^1)=\sqrt{(\sin^2\theta_{max} -  \sin^2 \theta)( \sin^2\theta - \sin^2\theta_{min})}.
\end{eqnarray}
\subsubsection{Conserved charges}
The energy of the nonrelativistic string configuration is given by,
\begin{eqnarray}
E_{NR}=\frac{2\sqrt{\mathfrak{g}}}{ \pi}\int_{\theta_{min}}^{\theta_{max}}d\theta \frac{(\sin^2\theta_{max} -  \sin^2 \theta)( \sin^2\theta - \sin^2\theta_{min})+\sin^2\theta \cos^2\theta}{\sqrt{(\sin^2\theta_{max} -  \sin^2 \theta)( \sin^2\theta - \sin^2\theta_{min})}}
\label{e27}
\end{eqnarray}
where the limits of the integral (\ref{e27}) are set to be,

On the other hand, the R- charges are given by
\begin{eqnarray}
J_1 =-\frac{\sqrt{\mathfrak{g}}}{ 2\pi}\int_{\theta_{min}}^{\theta_{max}}d\theta \frac{\sin^2\theta }{\sqrt{(\sin^2\theta_{max} -  \sin^2 \theta)( \sin^2\theta - \sin^2\theta_{min})}},
\end{eqnarray}
\begin{eqnarray}
J_2 =-\frac{\sqrt{\mathfrak{g}}}{2 \pi}\int_{\theta_{min}}^{\theta_{max}}d\theta \frac{\sin^2\theta -\frac{2}{3}}{\sqrt{(\sin^2\theta_{max} -  \sin^2 \theta)( \sin^2\theta - \sin^2\theta_{min})}}.
\end{eqnarray}

Finally, the angular difference between the end points of the solition is given by,
\begin{eqnarray}
\Delta \phi = 2\int_{\theta_{min}}^{\theta_{max}}d\theta \frac{1 }{\sqrt{(\sin^2\theta_{max} -  \sin^2 \theta)( \sin^2\theta - \sin^2\theta_{min})}} +\phi_{ct}.
\label{e30}
\end{eqnarray}

In order to evaluate the above integral (\ref{e30}) we set the upper limit to be $ \theta_{max}=\frac{\pi}{2} $. On top of that, we consider the excitation of the string soliton to be highly \emph{localized} along the equatorial plane of $ S^3 $ which thereby sets the lower limit of the integral as $ \theta_{min}\sim \frac{\pi}{2}-\epsilon $ with $ \epsilon \ll 1$. With these inputs, the divergences appearing in (\ref{e30}) can be cured by adding a suitable counter term of the form,
\begin{eqnarray}
 \phi_{ct} = \frac{\pi }{2\epsilon} -\frac{2}{\cos\theta_{min}}\tanh^{-1}(1-\epsilon).
\end{eqnarray}

This finally yields the \emph{regularised} angular difference,
\begin{eqnarray}
\Delta \phi^{(reg)} \sim \Delta\varphi_m  \approx \frac{\pi}{6} \epsilon + \mathcal{O}(\epsilon^3)
\end{eqnarray}
such that the product $ \sqrt{\mathfrak{g}}\Delta\varphi_m $ ($ \gg 1 $) is always \emph{finite} in the strong coupling ($ \mathfrak{g}\rightarrow \infty $) limit of the dual SMT. We apply similar arguments for the rest of the paper. 
\subsubsection{The spectrum}
The conserved charges are to be evaluated as,
\begin{eqnarray}
\label{e34}
E_{NR}&=&\frac{2\sqrt{\mathfrak{g}}}{ \pi}\int_{\theta_{min}}^{\pi/2}d\theta \frac{\cos\theta(2\sin^2\theta - \sin^2\theta_{min})}{\sqrt{ \sin^2\theta - \sin^2\theta_{min}}},\\
J_{1} &= &-\frac{\sqrt{\mathfrak{g}}}{2 \pi}\int_{\theta_{min}}^{\pi/2}d\theta \frac{\sin^2\theta }{\cos\theta \sqrt{\sin^2\theta - \sin^2\theta_{min}}},\\
J_{2} &=& -\frac{\sqrt{\mathfrak{g}}}{2 \pi}\int_{\theta_{min}}^{\pi/2}d\theta \frac{\sin^2\theta -\frac{2}{3}}{\cos\theta \sqrt{\sin^2\theta - \sin^2\theta_{min}}}.
\label{e36}
\end{eqnarray}

Setting, $ \theta_{min}\sim \frac{\pi}{2}- \epsilon $ as before we find,
\begin{eqnarray}
\label{e40}
E_{NR} &\sim & \frac{2\sqrt{\mathfrak{g}}}{ \pi} \epsilon + \mathcal{O}(\epsilon^3)\\
J_1 & \sim & - \frac{\sqrt{\mathfrak{g}}}{ 4\pi}\Delta \varphi_{m}+\frac{\sqrt{\mathfrak{g}}}{ 2\pi}(\epsilon + \frac{\epsilon ^3}{6}+\cdots)\\
J_2 & \sim & - \frac{\sqrt{\mathfrak{g}}}{ 12\pi}\Delta \varphi_{m}+\frac{\sqrt{\mathfrak{g}}}{ 2\pi}(\epsilon + \frac{\epsilon ^3}{6}+\cdots).
\label{e42}
\end{eqnarray}

Combining (\ref{e40})-(\ref{e42}) we obtain the two spin nonrelativistic magnon spectrum,
\begin{eqnarray}
E_{NR}-J_1\sim 3J_2 +\frac{\sqrt{\mathfrak{g}}}{2 \pi}  \Delta \varphi_{m}   +\mathcal{O}(\epsilon^3).
\label{e43}
\end{eqnarray}
\subsubsection{A spin chain interpretation}
The purpose of this section is to revisit the SMT limit \cite{Harmark:2018cdl} of magnon bound states in $ \mathcal{N}=4 $ SYM and provide a spin chain interpretation of the nonrelativistic magnon spectrum obtained above in (\ref{e43}). The near BPS corner that we are talking about is obtained as a decoupling limit \cite{Harmark:2014mpa} of the sector spanned by \emph{heavy} single trace operators of the form $ \mathcal{O}_{J_1J_2}\sim tr (\Phi_{X}^{J_1}\Phi_Z^{J_2}) +\cdots$ in $ \mathcal{N}=4 $ SYM. 

We start with the relativistic dispersion relation corresponding to $ J_2 $ magnon bound states in $ \mathcal{N}=4 $ SYM \cite{Dorey:2006dq},
\begin{eqnarray}
\Delta - J_1 = \sqrt{J_2^2 + \frac{\lambda}{\pi^2}\sin^2\frac{p_m}{2}}.
\label{E46}
\end{eqnarray}

Considering the large magnon bound states ($ J_2 \gg \sqrt{\lambda} $), in the weak coupling regime of $ \mathcal{N}=4 $ SYM theory we obtain,
\begin{eqnarray}
\Delta - J_1 \approx J_2 + \frac{\lambda}{2 \pi^2 J_2}\sin^2\frac{p_m}{2}+\mathcal{O}(\lambda^2).
\label{E47}
\end{eqnarray}

Finally, considering the decoupling limit \cite{Harmark:2018cdl}, we arrive at the dispersion relation corresponding to two spin magnons in the dual SMT
\begin{eqnarray}
\Delta_{NR}-J_1 \sim J_2 + \frac{\mathfrak{g}p^2_m}{8 \pi^2 J_2}  + \cdots
\label{e32}
\end{eqnarray} 
where $ \mathfrak{g} $ is finite but small and $ p_m \ll 1 $.

Notice that, in the strong coupling limit ($ \mathfrak{g}\gg 1 $) the dual quantum mechanical (Spin-Matrix theory) degrees of freedom are strongly coupled. Therefore, like in any other nonperturbative quantum mechanical system, it is hard to carry out direct (analytic) computations for any of its associated charges ($ H $ or $ Q $). The only way one can make predictions about the strong coupling physics is based on the notion of holography which for the present case is the nonrelativistic (SMT) sigma model (\ref{e2}) that we are studying.

Therefore, considering holography as a basic guiding principle, we propose the following strong coupling expansion for the second R-charge,
\begin{eqnarray}
J_2 \sim \sqrt{\mathfrak{g}}f (p_m)\sim \frac{\sqrt{\mathfrak{g}}}{\pi} \ell_{1}(  \bar{p}_m +\ell_2 \bar{p}^3_m + \cdots)\sim \ell_1 \tilde{J}_2~;~\bar{p}_m = \frac{p_m}{c}\ll 1.
\label{e49}
\end{eqnarray}
Notice that the above expansion (\ref{e49}) is motivated from our string theory results in (\ref{e42}) where we use the holographic dictionary that relates the geometrical angular difference with the magnon momentum along the spin chain namely, $ \Delta\varphi_m \sim  \frac{p_m}{c} \sim \bar{p}_m$ \cite{Hofman:2006xt}. We claim that with (\ref{e49}), the dispersion relation (\ref{E46}) will coincide with the string theory results in (\ref{e43}) and this will automatically fix the constants in the expansion of the R-charge (\ref{e49}).

Using (\ref{e49}) and considering the strong coupling replacement $ \lambda \sim \frac{ \mu \mathfrak{g}}{c^2} $ \cite{Harmark:2018cdl} we find,
\begin{eqnarray}
\Delta_{NR}-J_1 \sim \ell_1 \tilde{J}_2 +\frac{\mu \sqrt{\mathfrak{g}}}{8 \pi \ell_1}\bar{p}_m + \cdots
\end{eqnarray}
which precisely matches with string theory results (\ref{e43}) subjected to the identifications, $ \ell_1 =3 $,  $ \mu = 12 $.
\subsection{Three spin magnons}
We now generalize our previous analyses by constructing a three spin solution. In the three spin case, we consider two of the spins are to be aligned along $ S^5 $ while the remaining one lies along the three sphere $ S^3 \subset AdS_5 $. From the perspective of the dual SMT, these string states should be realised as the near BPS limit \cite{Harmark:2014mpa} of the sector spanned by single trace operators of the form $ \mathcal{O}_{J_1 J_2 S}\sim tr (\Phi_X^{J_1}(n^{\mu}D_{\mu})^S \Phi_Z^{J_2})+\cdots $ in $ \mathcal{N}=4 $ SYM. 

Below, we propose a string embedding of the following form,
\begin{eqnarray}
t &=& \kappa \sigma^0 ~;~ \eta = \sigma^1 ~;~ \psi =0 ~;~ \rho = \rho (\sigma^1)~;~\varphi = \nu \sigma^0 +\sigma^1 \nonumber\\
 \theta_1 & =& \theta (\sigma^1) ~;~ \theta_2 =0 ~;~\phi_1 =\omega_{1} \sigma^0 + \sigma^1 ~;~\phi_2 = \omega_{2} \sigma^0 + \sigma^1
\label{e70}
\end{eqnarray}
which results in the sigma model Lagrangian of the following form,
\begin{eqnarray}
\mathcal{L}_{NG}=-\frac{\nu}{2}\tanh^2\rho +\cosh^{-2}\rho \left( \frac{1}{2}(\omega_1 + \omega_2 )\sin^2\theta - \frac{\omega_2}{3}\right)\nonumber\\
 +\frac{\kappa}{8}\sinh^2\rho
+\frac{\kappa}{2}\cosh^2\rho (\rho'^2 +\theta'^2 + \sin^2\theta \cos^2\theta)\nonumber\\
+\frac{\kappa}{2}\sinh^2\rho \left( \sin^2\theta - \frac{1}{3}\right)\left(\frac{2}{3}+ \sin^2\theta\right).
\end{eqnarray}
\subsubsection{Equations of motion}
The resulting equations of motion could be formally expressed as
\begin{eqnarray}
\nu \tanh\rho \cosh^{-2}\rho +2 \cosh^{-3}\rho \sinh\rho \left( \frac{1}{2}(\omega_1 + \omega_2 )\sin^2\theta - \frac{\omega_2}{3}\right)-\frac{\kappa}{4}\sinh\rho \cosh\rho \nonumber\\
+\kappa \cosh^2\rho \rho'' +\kappa \cosh\rho \sinh\rho \rho'^2 - \kappa \cosh\rho \sinh\rho (\theta'^2 +\sin^2\theta \cos^2\theta)\nonumber\\
- \kappa \sinh\rho \cosh\rho \left( \sin^2\theta - \frac{1}{3}\right)\left(\frac{2}{3}+ \sin^2\theta\right)=0,
\label{e72}
\end{eqnarray}
\begin{eqnarray}
(\omega_1 + \omega_2)\cosh^{-2}\rho \sin\theta \cos\theta - \kappa \cosh^2\rho \theta'' -2\kappa \cosh\rho \sinh\rho \rho' \theta'\nonumber\\
+\kappa \cosh^2\rho \sin\theta \cos\theta (1-2\sin^2\theta)+\kappa \sinh^2\rho\sin\theta \cos\theta \left(\frac{1}{3}+ 2\sin^2\theta\right)=0.
\label{e73}
\end{eqnarray}

In order to solve (\ref{e72}) and (\ref{e73}), like before we consider strings those are moving at a fixed radial distance $\rho = \rho_c$. Substituting this ansatz into (\ref{e72}) and considering the fact that strings are moving near the centre ($ \rho_c \ll 1 $) of $ AdS_5 $ one finds,
\begin{eqnarray}
\theta'^2 \approx  \beta_c \sin^2\theta - \chi_c
\label{e77}
\end{eqnarray}
where we define new entities,
\begin{eqnarray}
\beta_c &\approx & \frac{\omega_1 +\omega_2}{\kappa} -\frac{4}{3}\\
\chi_c &\approx &\frac{1}{36}-\left( \frac{\nu}{\kappa}-\frac{2\omega_2}{3\kappa}\right) .
\end{eqnarray}

Following similar steps, from (\ref{e73}) we obtain,
\begin{eqnarray}
\theta'^2 \approx - \sin^4\theta +\left( 1+\frac{\omega_1 +\omega_2}{\kappa}\right) \sin^2\theta - \mathcal{C} 
\label{e80}
\end{eqnarray}
where $ \mathcal{C} $ is the constant of integration.

Adding (\ref{e77}) and (\ref{e80}) together we find,
\begin{eqnarray}
\theta'  \approx \frac{1}{2}\sqrt{(\sin^2\theta_{max}-\sin^2\theta)(\sin^2\theta - \sin^2\theta_{min})}
\end{eqnarray}
where the following relations are satisfied,
\begin{eqnarray}
\sin^2\theta_{max}+\sin^2\theta_{min}&=& 2\beta_c +\frac{7}{3}\\
\sin^2\theta_{max}\sin^2\theta_{min}&=&\chi_c + \mathcal{C}.
\end{eqnarray}
\subsubsection{Conserved charges}
Below we note down the conserved charges associated with the sigma model. The energy associated with the sigma model is give by,
\begin{eqnarray}
E_{NR}\approx \frac{\sqrt{\mathfrak{g}}}{\pi}\int_{\theta_{min}}^{\theta_{max}}d\theta \frac{(\sin^2\theta_{max}-\sin^2\theta)(\sin^2\theta - \sin^2\theta_{min})+4\sin^2\theta \cos^2\theta}{\sqrt{(\sin^2\theta_{max}-\sin^2\theta)(\sin^2\theta - \sin^2\theta_{min})}}\nonumber\\
+\frac{\sqrt{\mathfrak{g}}}{\pi}\rho_c^2\int_{\theta_{min}}^{\theta_{max}}d\theta \frac{1+4 \left( \sin^2\theta - \frac{1}{3}\right)\left(\frac{2}{3}+ \sin^2\theta\right)}{\sqrt{(\sin^2\theta_{max}-\sin^2\theta)(\sin^2\theta - \sin^2\theta_{min})}}.
\end{eqnarray}

On the other hand, the R-charges are given by
\begin{eqnarray}
J_1 \approx - \frac{\sqrt{\mathfrak{g}}}{\pi}\int_{\theta_{min}}^{\theta_{max}}d\theta \frac{\sin^2\theta}{\sqrt{(\sin^2\theta_{max}-\sin^2\theta)(\sin^2\theta - \sin^2\theta_{min})}},
\end{eqnarray}
\begin{eqnarray}
J_2 \approx - \frac{\sqrt{\mathfrak{g}}}{\pi}\int_{\theta_{min}}^{\theta_{max}}d\theta \frac{\sin^2\theta - \frac{2}{3}}{\sqrt{(\sin^2\theta_{max}-\sin^2\theta)(\sin^2\theta - \sin^2\theta_{min})}}.
\end{eqnarray}

Finally, the spin angular momentum is noted down to be,
\begin{eqnarray}
S_{\varphi}\approx \frac{\sqrt{\mathfrak{g}}}{\pi}\rho^2_c \int_{\theta_{min}}^{\theta_{max}}\frac{d\theta}{\sqrt{(\sin^2\theta_{max}-\sin^2\theta)(\sin^2\theta - \sin^2\theta_{min})}}.
\end{eqnarray}
\subsubsection{The spectrum}
Like before, we set the limit in which the angular difference between the end points of the string soliton is infinitesimally small namely $ \theta_{min} \sim \frac{\pi}{2}-\epsilon$ where we set the upper limit of the integral as, $ \theta_{max}=\frac{\pi}{2} $. This yield the conserved entities in the following form,
\begin{eqnarray}
\label{e88}
E_{NR} &\approx & S_{\varphi}+ \frac{4\sqrt{\mathfrak{g}}}{\pi}\epsilon +\cdots \\
J_1 & \approx & -\frac{\sqrt{\mathfrak{g}}}{4 \pi}\Delta \varphi_{m}+\frac{\sqrt{\mathfrak{g}}}{ \pi}\epsilon +\cdots \\
J_2 & \approx & -\frac{\sqrt{\mathfrak{g}}}{12 \pi}\Delta \varphi_{m}+\frac{\sqrt{\mathfrak{g}}}{ \pi}\epsilon +\cdots \\
S_{\varphi} & \approx & \frac{\sqrt{\mathfrak{g}}}{4 \pi}\rho_c^2 \Delta \varphi_{m}
\label{e91}
\end{eqnarray}
where, $ \Delta \varphi_{m} \sim \frac{\pi}{3}\epsilon $ is the \emph{regularised} angular difference between the end points of the string soliton. Combining (\ref{e88})-(\ref{e91}), we finally arrive at the three spin nonrelativistic giant magnon spectrum at strong coupling,
\begin{eqnarray}
E_{NR} - J_1 \sim  S_{\varphi}+3J_2  +\frac{\sqrt{\mathfrak{g}}}{2\pi}\Delta \varphi_{m}+\cdots
\end{eqnarray}
which generalises our previous results on nonrelativistic two spin magnons in (\ref{e43}).
\section{Summary and final remarks}
To summarise, the present paper explores nonrelativistic multispin magnon excitations by probing Spin-Matrix dynamics of strings over $ U(1) $ Galilean geometry. Considering two spin SMT limit of strings on $R \times S^5$ the corresponding magnon bound states are obtained in the limit of strong coupling. We further generalize our results by constructing nonrelativistic magnon bound states corresponding to three spin excitations. In the stringy construction, two of the spins are considered to be aligned along $S^5$ while the remaining one is taken to be aligned along $S^3 \subset AdS_5$. Although we provide a quantum mechanical interpretation for the two spin spectrum, however, a similar interpretation for the three spin spectrum is not ready at the moment. Below, we outline possible steps to construct such dispersion relations from the perspective of an underlying spin chain model.

Notice that, the three spin solution that we construct in this paper corresponds to a bound state of $ J_2 $ nonrelativistic magnons that also carries a spin $ S \sim S_{\varphi} $.
Therefore, to validate these results from the perspective of an underlying spin-chain model, as a first step, one needs to construct an \emph{exact} (in t'Hooft coupling ($ \lambda $)) dispersion relation for $ J $ magnon bound states carrying a spin $ S $ in some supersymmetry representation of $ \mathcal{N}=4 $ SYM in the $ SU(1,1|2) $ sector. Here, $ J $ corresponds to one of the Cartan generators of the unbroken R- symmetry group $ SO(4)\sim SU(2)\times SU(2) $.  One of the possibilities to construct these magnon states is to attempt to find an $ S $ matrix and the associated Bethe equation for the  $ SU(1,1|2) $ sector. One can initiate this by considering the vacuum state in $ SU(1,1|2) $ sector and thereby turning on excitations along the spin chain. This leads to factorized scattering between these excitations/magnon states which eventually leads to asymptotic Bethe equations for the $ SU(1,1|2) $ sector. Looking at the roots of this Bethe equation one could finally identify the local charges associated with the spin chain model which will eventually lead to the three spin dispersion relation that we are looking for. It will be nice to investigate all these possibilities as a project in the future. 
\\\\{\bf {Acknowledgements :}}
 The author is indebted to the authorities of IIT Roorkee for their unconditional support towards researches in basic sciences. The author would like to acknowledge The Royal Society, UK for financial assistance. The author would also like to acknowledge the Grant (No. SRG/2020/000088) received from The Science and Engineering Research Board (SERB), India.

\end{document}